%% file: submission_main.tex
\documentclass[journal]{IEEEtran}
\usepackage{amsmath,amsfonts}
\usepackage{algorithmic}
\usepackage{algorithm}
\usepackage{array}
\usepackage[caption=false,font=normalsize,labelfont=sf,textfont=sf]{subfig}
\usepackage{textcomp}
\usepackage{stfloats}
\usepackage{url}
\usepackage{verbatim}
\usepackage{graphicx}
\usepackage{cite}
\usepackage{csquotes}
\usepackage{amssymb}
\usepackage{enumitem}
\usepackage{tikz}
\usepackage{amsmath}
\usetikzlibrary{arrows.meta, positioning, fit, backgrounds, calc, shapes.geometric}
\definecolor{myred}{RGB}{180, 0, 0}
\definecolor{highlightblue}{RGB}{255, 182, 182}

\tikzset{
  mainbox/.style={
    rectangle, rounded corners=8pt,
    draw=black, line width=1.5pt, fill=white,
    text width=5cm, align=center,
    inner sep=10pt, font=\sffamily
  },
  sidebox/.style={
    rectangle, rounded corners=6pt,
    draw=black, line width=1.2pt, fill=white,
    text width=4cm, align=center,
    inner sep=8pt, font=\sffamily\small
  },
  subbox/.style={
    rectangle, rounded corners=4pt,
    draw=black, line width=1pt, dashed,
    fill=highlightblue, text width=4.4cm, align=left,
    inner sep=8pt, font=\sffamily\small
  },
  outerbox/.style={
    rectangle, rounded corners=10pt,
    draw=black, line width=1.5pt, fill=white,
    inner sep=10pt
  },
  sogarrow/.style={-Stealth, line width=1.5pt, color=black}
}

\tikzset{
  bu1/.style={draw=none, fill=none, minimum width=1.05cm,
              minimum height=0.65cm, align=center, rectangle,
              font=\sffamily\small, text=black},
  bu2/.style={draw=none, fill=none, minimum width=1.05cm,
              minimum height=0.65cm, align=center, rectangle,
              font=\sffamily\small, text=black},
  mac/.style={draw=black, fill=none, minimum width=1.05cm,
              minimum height=0.65cm, align=center, rectangle,
              font=\sffamily\small},
  macswap/.style={draw=red!70!black, fill=none, minimum width=1.05cm,
                  minimum height=0.65cm, align=center, rectangle,
                  line width=1.2pt, font=\sffamily\small},
  noi/.style={draw=red!70!black, fill=none, minimum width=1.05cm,
              minimum height=0.65cm, align=center, rectangle,
              font=\sffamily\small},
  bit/.style={draw=none, fill=none, minimum width=1.05cm,
              minimum height=0.3cm, align=center,
              font=\sffamily\scriptsize, text=black!70},
  lbl/.style={align=left, text width=2.0cm, font=\sffamily\footnotesize},
  msarr/.style={->, black!70, line width=0.65pt},
  swp/.style={<->, red!75!black, line width=0.6pt, dashed}
}

\usepackage{tikz,stackengine,mathtools}
\usetikzlibrary{matrix,calc}
\newsavebox{\foobox}

\usepackage{tikz}
\usetikzlibrary{arrows,shapes.misc,chains,scopes}
\usepackage{pgfplotstable}
\pgfplotsset{compat=1.15}

\usepackage{physics}
\DeclarePairedDelimiter\aparen{\lparen}{\rparen}

\DeclarePairedDelimiter\abrace{\lbrace}{\rbrace}

\renewcommand{\pqty}[1]{\aparen*{#1}}

\renewcommand{\Bqty}[1]{\abrace*{#1}}

\newcommand{\giv}{\,\middle|\,}

\def\C{\mathbb{C}}

\newcommand{\nout}{n_{\mathrm{out}}}
\newcommand{\kout}{k_{\mathrm{out}}}
\newcommand{\ninn}{n_{\mathrm{inn}}}
\newcommand{\kinn}{k_{\mathrm{inn}}}

\DeclareMathOperator*{\argmin}{\arg\min}
\renewcommand{\hat}{\widehat}
\newcommand{\X}{\mathcal{X}}
\newcommand{\Xagg}{\X_{\text{agg}}}
\newcommand{\Y}{\mathbf{Y}}
\newcommand{\Xb}{\mathbf{X}}
\newcommand{\y}{\mathbf{y}}
\newcommand{\z}{\mathbf{z}}
\newcommand{\x}{\mathbf{x}}
\newcommand{\rb}{\mathbf{r}}

\newcommand{\Zb}{\mathbf{Z}}
\newcommand{\Rb}{\mathbf{R}}
\newcommand{\R}{\mathbb{R}}

\DeclareMathOperator*{\Prb}{\mathbb{P}}
\renewcommand{\Pr}[1]{\Prb\pqty{#1}}

\title{Soft-Output GRAND-Aided Macrosymbol}

\begin{document}

\author{
Sarah~Khalifeh,
Alexander~Mariona,
Ken~R.~Duffy,
and~Muriel M\'edard
\thanks{S. Khalifeh and K. R. Duffy are with Northeastern University, (e-mail: khalifeh.s@northeastern.edu; k.duffy@northeastern.edu).}
\thanks{A. Mariona and M. M{\'e}dard are with Massachusetts Institute of Technology, (e-mail: amariona@mit.edu; medard@mit.edu).}
\thanks{This material is based upon work supported by the National Science Foundation under Grant No. (ECCS-2433994 and ECCS-2433996). This work was supported by the Defense Advanced Research Projects Agency (DARPA) under Grant HR00112120008.}
}

\maketitle

\begin{abstract}
Guessing Random Additive Noise Decoding-Aided Macrosymbol (GRAND-AM) provides a
coding-based solution for non-orthogonal multiple access (NOMA) systems,
enabling joint multiuser detection and error correction. GRAND-AM outperforms prior methods in symbol error rates while providing
hard-detection output. We introduce soft-output GRAND-AM
(SOGRAND-AM), a macrosymbol-level joint multiuser detection and decoding
framework which generates calibrated bitwise a posteriori error probabilities. SOGRAND-AM extends GRAND-AM by enabling soft-decision outer forward error correction decoders, transparently integrating with existing decoders. SOGRAND-AM achieves approximately 2 dB improvement in bit error rate at the output of the
outer soft-input decoder compared to hard input with GRAND-AM.
\end{abstract}


\section{Introduction}
The recent shift from orthogonal multiple access (OMA) to non-orthogonal
multiple access (NOMA) is driven by ultra-reliable low-latency
applications, many-user Internet of Things (IoT) applications, and demands for
higher spectral efficiency \cite{global_conn,chen2018ultra,durisi2016toward}. Such systems require effective mitigation of multiple access interference (MAI) to meet target error rates\cite{saito2013non,dai2015non,7405722}.
Guessing Random Additive Noise Decoding-Aided Macrosymbol {(GRAND-AM)}
addresses MAI through joint multiuser detection (MUD) and error correction
using short inner MAC rate-splitting (IMACRS) codes, reformulating the
multiuser detection problem at the macrosymbol level and employing Symbol-level
ORBGRAND for near maximum-likelihood (ML) decoding without demapping
\cite{yang2023multiuser, yang2024nonorthogonal, yang2025grand, wei2023symbol,
Kizilates25ORBRGANDAI}. GRAND-AM has been shown to outperform time division multiple
access (TDMA) by 6 dB, even with
imperfect channel estimation \cite{yang2024nonorthogonal}. GRAND-AM provides hard-decision MAC output and so is exclusively compatible with hard-decision outer decoders. Advanced
forward error correction (FEC) decoders, however, take advantage of soft-input in the form
of log-likelihood ratios (LLRs) to improve decoding\cite{goldsmith2005wireless}.

\begin{figure*}[t]
\centering
\hspace{-1cm}
\includegraphics[width=0.95\linewidth, keepaspectratio]{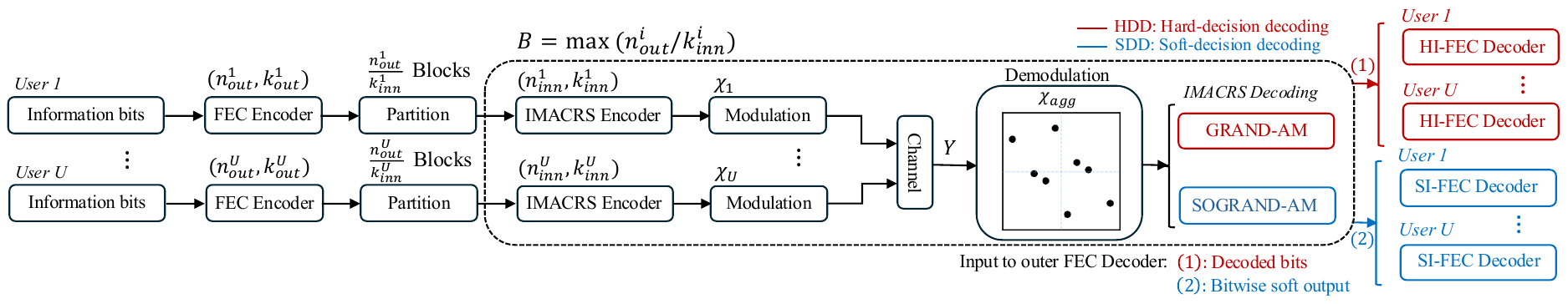}
\vspace{-0.3cm}
\caption{A $U$-user NOMA system where the receiver
performs joint demodulation over $\mathcal{X}_{\text{agg}}$ and applies:
GRAND-AM, which feeds hard-decided bits to the hard-input FEC (HI-FEC) decoders; or
the proposed SOGRAND-AM, which feeds calibrated bitwise SO to the soft-input FEC (SI-FEC) decoders.}
\label{fig:HDD-SDD}
\vspace{-0.5cm}
\end{figure*}

We introduce Soft-Output GRAND-AM (SOGRAND-AM), an extension of GRAND-AM that
provides LLR output from the MAC layer. SOGRAND-AM adapts the soft-output
computation techniques introduced by SOGRAND, a soft-input soft-output channel
decoder that accurately estimates the posterior probability that its decoding
is correct\cite{yuan2025soft}. By producing per-user blockwise and bitwise soft-output, SOGRAND-AM integrates transparently with existing soft-input FEC decoders and codes. The NOMA system illustrated in Fig.~\ref{fig:HDD-SDD}, contrasts
hard-decision decoding (HDD) and soft-decision
decoding (SDD) of the FEC codes using GRAND-AM and SOGRAND-AM for IMACRS decoding, respectively.

Although recent work has considered GRAND as a NOMA solution, the
resulting decoders are based on GRAND-AM variations with less sophisticated guesswork
mechanisms. The MOSS-GRAND decoder \cite{MOSSGRAND} is a joint MUD and decoding
scheme that performs guesswork over the joint symbols, like GRAND-AM. Unlike
GRAND-AM, which preceded it, MOSS-GRAND does not use Symbol-level ORBGRAND, but
instead guesses joint symbols in an ordering induced by the $L_1$ norm, using a
local repair scheme to enforce joint membership. Zor et
al.\cite{zor20256gdownlinknomacrcaided} propose integrating CRC-aided GRAND
with successive interference cancellation (SIC) in a downlink power-domain NOMA
framework and use GRAND to mitigate error propagation during interference
cancellation. More recently, an iterative joint detection and ORBGRAND framework has been proposed for massive MU-MIMO systems. Instead of performing guesswork over macrosymbols, the receiver alternates between LMMSE detection, ORBGRAND decoding, and adaptive interference cancellation\cite{11614848}.

The remainder of the paper is structured as follows. Sections
\ref{sec:system-model} and \ref{sec:grand-am} present the system model and some background on GRAND-AM. Section~\ref{sec:soft-output} details how
SOGRAND-AM computes soft-output for the MAC layer.
Section~\ref{sec:performance-evaluation} presents empirical results
demonstrating that SOGRAND-AM outperforms both GRAND-AM and other
state-of-the-art NOMA solutions. We conclude in Section~\ref{sec:discussion}.

\section{System Model}\label{sec:system-model}

We consider a $U$-user multiple-access channel (MAC) with each user employing
a two-layer coding scheme: an outer layer forward
error correction (FEC) code and an inner layer inner MAC
rate-splitting (IMACRS) code. The FEC code provides redundancy for reliable transmission in the presence of noise,
while the IMACRS code guards against inter-user interference. Throughout, random variables are denoted by uppercase and their realizations by lowercase. Constants are written in both cases, with the distinction being clear from context. Vector quantities are written in boldface and indexed by
subscripts. Superscripts are used with both scalar and vector quantities to
specify properties such as to which user the quantity corresponds. We let $[n]:=\{1, 2, \dots, n\}$.

Each user encodes and modulates their transmission independently (Fig.~\ref{fig:HDD-SDD}). User $i$ employs an $[\nout^i,\kout^i]$ FEC code and
a $[\ninn^i,\kinn^i]$ IMACRS code, such that $\kinn^i$ divides $\nout^i$. The
FEC code maps $\kout^i$ information bits to produce $\nout^i$ coded
bits. These bits are partitioned into $\nout^i / \kinn^i$ blocks of $\kinn^i$ bits, each encoded using the IMACRS code to produce
$\ninn^i$ coded bits per block. User $i$ modulates each block using a discrete,
complex constellation of size $m^i$, yielding $s=\ninn^i/\log_2(m^i)$
transmitted symbols per IMACRS codeword assuming $\log_2(m^i)$ divides $\ninn^i$.
Although the code parameters and $m^i$ may vary between users, we
assume that $s$ is constant for ease of exposition. The per-user transmitted
symbols combine additively to form a \emph{macrosymbol}.
We assume a memoryless additive-noise fading channel. The
fading coefficients are independently sampled from a continuous distribution
with each transmission, and this distribution does not vary over time.
Similarly, the channel noise is sampled independently from a static continuous
distribution. The channel output is $ Y = N + \sum_{i=1}^U H^i X^i$, where $X^i\in\C$ is the symbol transmitted by user $i$, $H^i\in\C$ is the user's channel gain, and $N\in\C$ is the channel noise. The receiver is
aware of all codes, constellations, channel gains, and the noise distribution.

\section{Joint Detection and Decoding with GRAND-AM}\label{sec:grand-am}

GRAND-AM reformulates the multiuser detection problem as a single-user
detection problem over an aggregate constellation. Denoting the constellation
of user $i$ by $\mathcal{X}_i \subset \mathbb{C}$, the set of possible
macrosymbols for a particular transmission is
\begin{equation}\label{eq:xagg}
    \Xagg = \Bqty{
        \sum_{i=1}^U h^i x^i \giv (x^1,\dots,x^U) \in
        \mathcal{X}_1 \times \dots \times \mathcal{X}_U
    }.
\end{equation}
Note that $\Xagg$ is defined in terms of the realized channel gains; since these gains are continuously distributed, each
macrosymbol almost surely corresponds to a unique per-user symbol set.
While \eqref{eq:xagg} describes the constellation for a particular transmission,
we abuse notation slightly and write $\Xagg^s$ for the set of length-$s$ macrosymbol
sequences, with each Cartesian factor taken from its
respective transmission.
GRAND-AM produces a soft-decision decoding for the IMACRS code by performing joint detection and error correction. Let $\y$ be a realization
of the received sequence $\Y\in\C^s$. We denote by $\hat{\y}\in\Xagg^s$ the
macrosymbol sequence which minimizes the coordinate-wise Euclidean distance to
$\y$. This is the jointly-optimal ML detection for the
additive white Gaussian noise (AWGN) channel \cite{yang2024nonorthogonal}. We
refer to $\hat{\y}$ as the \emph{hard-detection sequence}. Explicitly,
\begin{equation}\label{eq:detect}
    \hat{\y}_j = \argmin_{x\in \Xagg}
    \lvert \y_j - x \rvert^2, \quad j\in[s].
\end{equation}

After computing the hard-detection, GRAND-AM performs joint error correction
using the IMACRS codes by employing Symbol-level ORBGRAND \cite{wei2023symbol},
an approximate ML soft-decision channel decoder which performs symbol
substitutions to find a decoding. A symbol
substitution is a pair $(x,j)$, where $x\in \Xagg\setminus\hat{\y}_j$ is the
macrosymbol to substitute at index $j$ of the hard-detection. Symbol-level
ORBGRAND guesses substitution combinations in approximately decreasing order of
the probability that $\Y=\y$ given that the channel input was $\hat{\y}$ with the
putative substitutions. To produce this guessing order, we first rank-order the
set of symbol substitutions by \emph{exceedance distance}, $ E(x,j) = \lvert \y_j - x \rvert^2
          - \lvert \y_j - \hat{\y}_j \rvert^2.$ By the detection rule \eqref{eq:detect}, $E(x,j)>0$ for all substitutions,
since we do not consider $x = \hat{\y}_j$ to be a valid substitution.

A combination of macrosymbol substitutions can be specified by a binary
sequence $\z$ of length $\pqty{\abs{\Xagg} - 1}s$, with $\z_k$
indicating whether the combination includes the substitution with rank-$k$
exceedance distance. Symbol-level ORBGRAND guesses combinations in
order of increasing \emph{logistic weight}, which is given by $ w_{\mathrm{L}}(\z) = \sum_{k:\, \z_k=1} k.$ For each $\z$, the corresponding substitutions are applied to $\hat{\y}$ to
produce a putative macrosymbol sequence $\x \in \mathbb{C}^s$.
If a combination performs multiple substitutions on the same index, it is considered invalid and skipped.

GRAND-AM employs symbol-level ORBGRAND to find a macrosymbol sequence for which every user's symbol sequence is simultaneously a valid modulated
IMACRS codeword. This is joint error correction: rather than decoding each user independently, the macrosymbol-level guessing process
maximizes the joint probability of all the users' decodings. Each user then independently decodes the outer FEC code.
As GRAND-AM does not produce soft output for the IMACRS decodings, only hard-decision FEC decoders are supported.

\section{Soft-Output GRAND-AM}\label{sec:soft-output}

SOGRAND-AM extends the GRAND-AM algorithm by providing soft-output for each
user's IMACRS decoding, enabling soft-decision decoding of the FEC codes. SOGRAND-AM adapts prior work on SOGRAND
\cite{Kizilates26SOGRAND,yuan2025soft} to the joint macrosymbol guesswork
process of GRAND-AM.

We first describe the single-user SOGRAND soft-output computation. The
transmitter uses an $[n,k]$ binary code and sends a uniformly-distributed
random codeword $\Xb\in\{0,1\}^n$ over an additive-noise channel. The receiver
observes a realization $\y$ of the channel output $\Y\in\{0,1\}^n$ along with a
realization $\rb$ of the reliability information $\Rb\in\mathbb{R}^n$. 

SOGRAND attempts to identify the noise effect $\Zb=\Xb \oplus \Y$, where $\oplus$ denotes binary addition, by rank-ordering noise effects
$\z\in\Bqty{0,1}^n$ in decreasing order of posterior probability $\Pr{\Zb=\z \giv \rb}$. SOGRAND then tests each putative transmission $\y\oplus\z$ for codebook membership. SOGRAND returns the first codeword found, or, for list decoding of length $L$, the first $L$ codewords found.

Given $\y$ and $\rb$, let $\z^j$ denote guess $j\in[2^n]$. For a list decoding $\mathcal{L}\subset \Bqty{0,1}^n$ of length $L$, let
$\mu(l)$ denote the guess at which list element $l\in[L]$ is identified. Therefore, the $l^{\text{th}}$ codeword in $\mathcal{L}$ is $\x^l = \y \oplus \z^{\mu(l)}.$ If the code has been chosen uniformly at random, the posterior probability that
$\Xb = \x^l$ is approximately \cite{yuan2025soft}
\begin{equation}\label{eq:single-so}
    \Pr{\Xb=\x^l \giv \rb} \approx
    \frac{
        \Pr{\Zb = \z^{\mu(l)} \giv \rb}
    }{
        P + Q \pqty{2^{k-n}}
    },
\end{equation}
where $P$ is the total probability of all noise effects corresponding to
identified codewords and $Q$ is the total probability of all unguessed noise
effects, i.e., $P = \sum_{l=1}^L \Pr{\Zb = \z^{\mu(l)} \giv \rb}$ and $ \quad
    Q = 1 - \sum_{j=1}^{\mu(L)}
        \Pr{\Zb = \z^{j} \giv \rb}.$
For brevity, we denote the conditioning event $\Rb=\rb$ by simply writing
$\rb$. Furthermore, the posterior probability that the list does not contain
the transmitted codeword is $\Pr{\Xb\notin \mathcal{L} \giv \rb} = 1-\sum_{l=1}^L \Pr{\Xb=\x^l \giv \rb}.$
To compute these approximations, SOGRAND maintains a running sum of the
probability of each guess, along with the individual probability of guesses
corresponding to codewords.

The \emph{blockwise soft-output} in \eqref{eq:single-so} denotes the
probability that a particular decoding is correct. The \emph{bitwise soft-output} $\mathbf{V}\in\R^n$, the posterior LLRs given a list decoding $\mathcal{L}$, can be
approximated as \cite{yuan2025soft}
\begin{align*}\label{eq:single-so-bit}
    \mathbf{V}_g &\approx \log\frac{
        \Pr{\Xb\notin \mathcal{L} \giv \rb} q^0_g +
        \sum_{\x\in\mathcal{L}:\, \x_g=0} \Pr{\Xb=\x \giv \rb}
    }{
        \Pr{\Xb\notin \mathcal{L} \giv \rb} q^1_g +
        \sum_{\x\in\mathcal{L}:\, \x_g=1} \Pr{\Xb=\x \giv \rb}
    }, \\
    q^b_g &= \Pr{\Xb_g=b \giv \rb_g}, \quad b\in\Bqty{0,1},\; g\in[n].
\end{align*}

\input{macrosymbol}

The goal of SOGRAND-AM is to compute analogous approximations for the
per-user IMACRS decodings. The
key difference from SOGRAND is that SOGRAND-AM terminates
only when a macrosymbol sequence satisfies all users’
IMACRS codebook membership checks simultaneously. During joint macrosymbol-level decoding,
SOGRAND-AM must perform bookkeeping similar to SOGRAND, but for every user
simultaneously, subject to a few key distinctions. First, although
SOGRAND-AM does not list-decode the IMACRS codes, the joint macrosymbol
guessing process may identify multiple candidate codewords for a given user
before finding a substitution that satisfies all users' IMACRS codes, with
each additional codeword identified updating the posterior probability that
the ultimate decoding is correct. Second, while every SOGRAND guess checks a
new putative transmission for codebook membership, a macrosymbol
substitution may leave a given user's symbol sequence unchanged, or map it
to a sequence already checked (Fig.~\ref{fig:macroswap}); a guess only
affects a user's posterior decoding probability if it corresponds to a
previously unexamined symbol sequence. As a result, different users may
identify different numbers of valid codewords, and the number of unique
code checks per user can differ.

To address these differences and correctly apply \eqref{eq:single-so}, we analyze the joint
process as a collection of single-user processes producing independent list
decodings for their IMACRS codes. Formally, let $\Xb\in\Xagg^s$ denote the
transmitted sequence formed by sampling a codeword $\Xb^i\in \X_i^s$ from each
user's IMACRS code uniformly at random. The receiver observes the channel
output $\Y\in\C^s$ and computes the hard-detection $\hat{\Y}\in\Xagg^s$, along
with the resulting exceedance distances. Let
$\Zb\in\Bqty{0,1}^{\pqty{\abs{\Xagg}-1}s}$ denote the substitution combination,
as described in Sec.~\ref{sec:grand-am}, such that applying $\Zb$ to $\hat{\Y}$
recovers $\Xb$. In the same way that the joint transmission $\Xb$ corresponds
to the per-user transmissions $\Xb^i$ for $i\in[U]$, the joint hard-detection
corresponds to the per-user hard-detections $\hat{\Y}^i \in \X^s_i$ and the
joint substitution $\Zb$ corresponds to the per-user substitutions
$\Zb^i\in\Bqty{0,1}^{\pqty{\abs{\X_i}-1}s}$.

Given a realization $\hat{\y}$ of $\hat{\Y}$, let
$\z^{j}\in\Bqty{0,1}^{\pqty{\abs{\Xagg}-1}s}$ denote the $j$th joint guess
which SOGRAND-AM will make.
We denote by $\z^{j,i}$ the corresponding per-user substitutions. Applying
$\z^{j}$ to $\hat{\y}$ yields a putative transmission $\x^{j}\in\Xagg^s$ which
uniquely corresponds to the per-user sequences $\x^{j,i}\in\X_i^s$ for all
$i\in[U]$. If $\x^{j',i} \neq \x^{j,i}$ for all $j' < j$, then $\x^{j,i}$ is a
new guess for user $i$, which we indicate by $\lambda^{j,i}\in\Bqty{0,1}$. Let
$T$ denote the number of joint guesses made before a joint decoding is found.
Thus, the final decoding for user $i$ is $\x^{T,i}$. Let $L^i$ denote the
number of codewords identified by user $i$ and let $\mu^i(l)\in[T]$ denote the
joint guess at which codeword $l\in[L^i]$ is first identified. It is possible
that $\mu^i(l) < T$ for all $l\in[L^i]$, namely, if $\x^{T,i}$ was not a new
guess.

As the receiver is aware of the channel model and constellations, SOGRAND-AM is
able to compute for each guess and each user the probability $\Pr{\Zb^i =
\z^{j,i} \giv \y}$, denoting the event $\Y=\y$ by simply writing $\y$. Assuming
that each IMACRS code has been chosen independently and uniformly at random,
the posterior probability that $\Xb^i = \x^{T,i}$ is approximately
\begin{equation}\label{eq:joint-so}
     \Pr{\Xb^i = \x^{T,i} \giv \y} \approx
    \frac{ \Pr{\Zb^i = \z^{T,i} \giv \y} }
    { P^i + Q^i \pqty{2^{\kinn^i-\ninn^i}}},
\end{equation}
where $P^i$ is the total probability of the per-user substitutions
corresponding to identified codewords and $Q^i$ is the total probability of
all unguessed per-user substitutions: $ P^i =
    \sum_{l=1}^{L^i} \Pr{\Zb^i = \z^{\mu^i(l),i} \giv \y}$ and $Q^i = 1 - \sum_{j=1}^{T} \lambda^{j,i} \Pr{\Zb^i = \z^{j,i} \giv \y}$. The posterior probability that the list of codewords $\mathcal{L}^i$ does not contain $\Xb^{i}$ is $ \Pr{\Xb^{i}\notin \mathcal{L}^i \giv \y} =
    1 - \sum_{l=1}^{L^i} \Pr{\Xb^i = \x^{\mu^i(l),i} \giv \y}.$
\begin{figure}
\vspace{-0.3cm}
    \centering
    \includegraphics[width=0.9\linewidth]{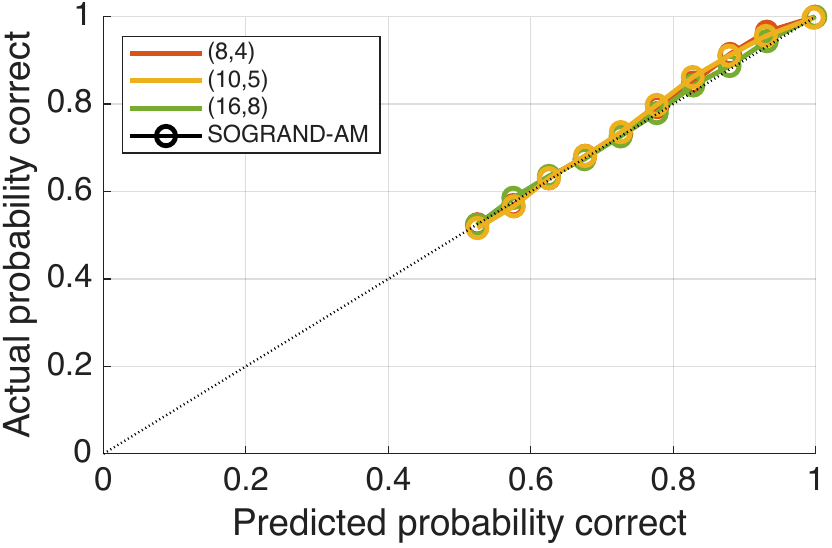}
    \caption{Calibration curve for the bitwise SO generated by SOGRAND-AM for User~1 (Strong user). Transmission is over an AWGN channel for different $(n^i_{\mathrm{inn}}, k^i_{\mathrm{inn}})$ at a BLER$\approx10^{-3}.$}
    \label{fig:bitwise_so}
    \vspace{-0.5cm}
\end{figure}
Eq. \eqref{eq:joint-so} gives the per-user blockwise soft-output that
SOGRAND-AM computes for the IMACRS decodings; as in the single-user case,
the per-user bitwise soft-output follows by marginalization. Denoting the binary demodulations of symbol sequences by an overline, the bitwise soft-output for bit $g\in[\ninn^i]$ of the IMACRS
decoding for user $i$ is approximately
\begin{align}\label{eq:bitwise-AM}
    \mathbf{V}^i_g &\approx \log \frac{
        \Pr{\Xb^i \notin \mathcal{L}^i} q^{i,0}_g +
        \sum_{\substack{\x\in\mathcal{L}^i:\\\overline{\x}_g=0}} \Pr{\Xb^i=\x\giv \y}
    }{
        \Pr{\Xb^i \notin \mathcal{L}^i} q^{i,1}_g +
        \sum_{\substack{\x\in\mathcal{L}^i:\\\overline{\x}_g=1}} \Pr{\Xb^i=\x\giv \y}
    },
\end{align}
with $q^{i,b}_g = \Pr{\overline{\Xb}^i_g = b \giv \y}$ and $b\in\Bqty{0,1}$.

SOGRAND-AM augments the single-user bookkeeping of SOGRAND with per-user tracking of actual substitutions, ensuring only
unique guesses are counted via $\lambda^{j,i}$. Circuit and in-silicon implementations of ORBGRAND demonstrate that large numbers of
codebook queries can be made in parallel, exceeding the numbers we will need for this application~\cite{abbas2021orbgrand,condo2021fixed,Kizilates25ORBRGANDAI,ji2025efficient}.
Given the independence of IMACRS encodings and decodings, the $B$ blocks can be processed in parallel (Fig.\ref{fig:HDD-SDD}), enabling low-latency decoding. The aggregate constellation $\Xagg$ grows
exponentially with the number of users $U$ and with the per-user
constellation size $m^i$. 

Nonetheless, the empirical query complexity
of SOGRAND-AM remains manageable in practice. At high
SNR, SOGRAND-AM requires fewer queries on average than the combined
per-user baseline, since each macrosymbol substitution increasingly
yields a unique per-user substitution (Fig.~\ref{fig:complexity_eq}). The reduction in average query number highlights SOGRAND-AM's modest complexity compared to FEC decoders.
Complexity could be further controlled via nearest-neighbor restriction for macrosymbol-likelihood computations, which limits the search space, and query thresholding in low-SNR
regimes~\cite{wei2023symbol,duffy2022ordered}.

\section{Performance Evaluation}
\label{sec:performance-evaluation}

 This section examines the accuracy of the bitwise SO $\mathbf{V}^i_g$ from~\eqref{eq:bitwise-AM} and quantifies the performance
gains from SOGRAND-AM. We consider a two-user MAC where each employs a CRC IMACRS code with a $(32,26)$ eBCH outer FEC code, which supports both ML hard- and near ML-soft-decision decoding. HI-GRAND is the HI-FEC (hard-input) decoder and ORBGRAND~\cite{duffy2022ordered} is the
SI-FEC (soft-input) decoder. To consider conditions favorable to successive interference cancellation (SIC), we introduce a $10$~dB SNR difference between User~1 (strong) and User~2 (weak).

At the output of the IMACRS decoder, we consider the bitwise SO generated by SOGRAND-AM. If the SO is accurate, then the calibration curve would follow the line $y=x$, as the predicted BER and the empirical BER would match. We find that SOGRAND-AM produces well-calibrated SO for multiple $(n^i_{\mathrm{inn}}, k^i_{\mathrm{inn}})$ IMACRS
codes (Fig.~\ref{fig:bitwise_so}).

\begin{figure}
    \centering
    \vspace{-0.4cm}
\includegraphics[width=0.85\linewidth]{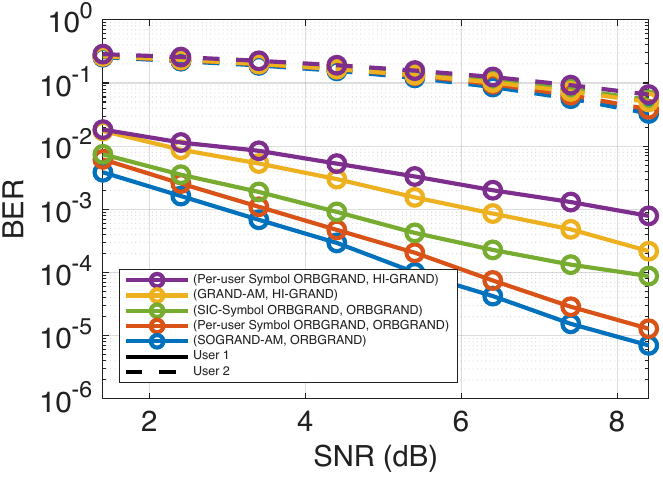}
    \caption{SI-FEC decoder BER for the $(32,26)$ eBCH code,
with users transmitting with a 10 dB SNR difference over a Rayleigh fading channel.\vspace{-0.4cm}}
    \label{fig:BER-unequal}
\end{figure}

We evaluate the BER at the output of the SI-FEC decoder for different (IMACRS,~FEC) decoding
pairs.
We consider
an $(8,4)$ CRC IMACRS with a complex BPSK constellation $\mathcal{X}^1=\mathcal{X}^2$, transmitting over a Rayleigh fading channel with $h_i\sim\mathcal{N}_C(0,1)$. SOGRAND-AM
maintains its performance advantage over both per-user Symbol-ORBGRAND and
SIC-Symbol ORBGRAND for both users, demonstrating robustness to power-asymmetric NOMA
conditions (Fig.~\ref{fig:BER-unequal}). A gap of roughly 2-3~dB between the hard- and soft-decision GRAND-AM variants is observed across the SNR range.

\begin{figure}
    \centering
\includegraphics[width=0.85\linewidth]{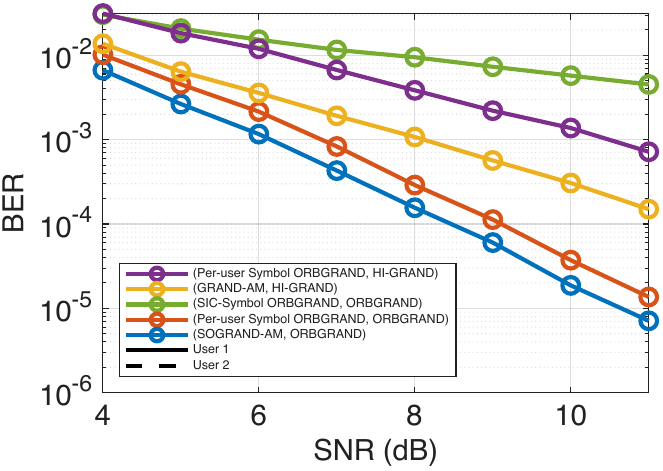}
    \caption{SI-FEC decoder BER for the $(32,26)$ eBCH code,
with users transmitting at equal power over a Rayleigh fading channel.}
    \label{fig:BER-equal-power}
    \vspace{-0.6cm}
\end{figure}

We consider the case where both users transmit using the same power. At a BER of $10^{-3}$,
the gap between (Per-user Symbol ORBGRAND, HI-GRAND) and (GRAND-AM,
HI-GRAND) is approximately 2~dB, reflecting the gain from joint macrosymbol
detection alone with HI-FEC decoding at both stages (Fig.~\ref{fig:BER-equal-power}). An
approximate 2~dB gain is observed between (GRAND-AM, HI-GRAND) and
(SOGRAND-AM, ORBGRAND) at the same BER level, attributable to the
transition from hard- to soft-decision FEC decoding enabled by the calibrated
SO produced by SOGRAND-AM. This gap slightly exceeds 2~dB at lower BER
($10^{-4}$).

\begin{figure}
\vspace{-0.45cm}
    \centering
    \includegraphics[width=0.85\linewidth]{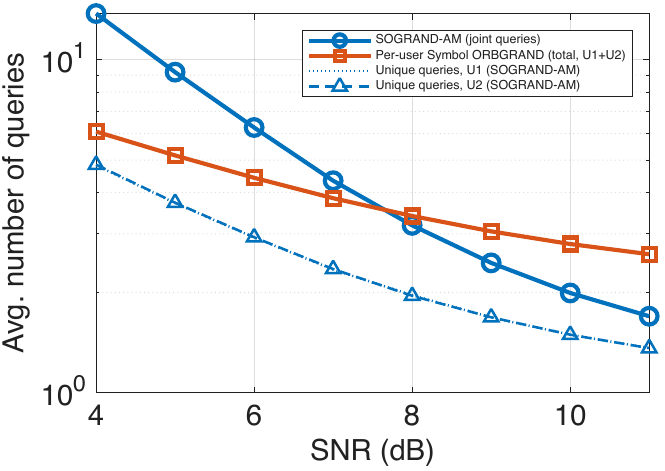}
    \caption{Average number of queries for a 2-user MAC with $(8,4)$ CRC IMACRS and
$(32,26)$ eBCH codes, equal-power users, Rayleigh fading. IMACRS decoding:
SOGRAND-AM and Per-user Symbol ORBGRAND.}
    \label{fig:complexity_eq}
    \vspace{-0.35cm}
\end{figure}

 For comparison, we report the sum of per-user query counts for Symbol-ORBGRAND for the equal-power setting, since it runs two
independent decoders (Fig.~\ref{fig:complexity_eq}). Overall, SOGRAND-AM yields better performance compared to per-user
symbol-ORBGRAND at the cost of a limited increase in joint decoding complexity at low SNR, while becoming more query-efficient as SNR
increases.

\begin{figure}
    \centering
    \vspace{-0.1cm}
    \includegraphics[width=0.85\linewidth]{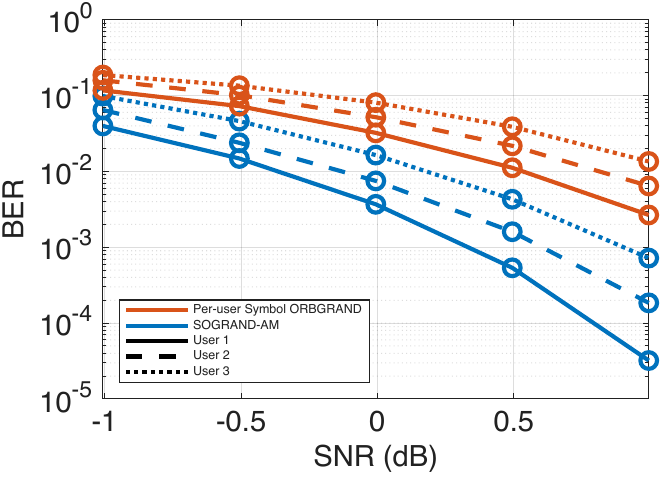}
    \caption{SI-FEC decoder BER for the $(384,192)$ 5G NR LDPC code under a power-imbalanced scenario (User~1 SNR advantage: 0.25~dB over User~2, 0.5~dB
    over User~3) over a Rayleigh fading channel.}
    \label{fig:ldpc_3user}
    \vspace{-0.5cm}
\end{figure}

We consider a three-user MAC with $(384,192)$ 5G NR LDPC outer FEC code \cite{richardson2018design} decoded using norm-min-sum decoding (Fig.~\ref{fig:ldpc_3user}). For all users, SOGRAND-AM consistently outperforms per-user Symbol-ORBGRAND. This shows that SOGRAND-AM naturally extends to larger $U$ while maintaining compatibility with conventional SI-FEC decoders.

\section{Discussion}\label{sec:discussion}
We introduce SOGRAND-AM, an extension of GRAND-AM, which provides a soft-output MAC module by tracking the likelihoods of the noise
effects queried by each user. SOGRAND-AM provides calibrated SO for
transparent use by any SI-FEC decoder without modifying the
code construction or outer decoder architecture. This calibrated soft-output leads to multiple-dB BER improvements at
the output of the FEC decoder compared to GRAND-AM and other NOMA solutions. These results demonstrate the
potential for SOGRAND-AM integration with a wide range of outer FEC
codes, including LDPC codes used in the 5G standard.
\bibliographystyle{IEEEtran}
\bibliography{references_am}
\end{document}

%% file: macrosymbol.tex
\begin{figure}[t]
\centering
\resizebox{\linewidth}{!}{
\begin{tikzpicture}[>=Stealth, line width=0.8pt]

\node[lbl] at (0,0) (Lhy)
    {Hard-detection\\$\mathbf{\hat{y}} \in \mathcal{X}_{\mathrm{agg}}^s$};

\node[mac,     right=0.2cm of Lhy]  (hy1){$a_1$};
\node[mac,     right=0.08cm of hy1] (hy2){$a_2$};
\node[macswap, right=0.08cm of hy2] (hy3){$a_1$};
\node[mac,     right=0.08cm of hy3] (hy4){$a_3$};
\node[mac,     right=0.08cm of hy4] (hy5){$a_2$};
\node[mac,     right=0.08cm of hy5] (hy6){$a_4$};
\node[mac,     right=0.08cm of hy6] (hy7){$a_3$};
\node[mac,     right=0.08cm of hy7] (hy8){$a_2$};

\node[bit, below=0.02cm of hy1] (hb1) {$(0,\,0)$};
\node[left=0.15cm of hb1, text=black!60, font=\sffamily\scriptsize,
      anchor=east, align=left]{};
\node[bit, below=0.02cm of hy2]{$(1,\,0)$};
\node[bit, below=0.02cm of hy3] (hy3bit){\textcolor{red!70!black}{$(0,\,0)$}};
\node[bit, below=0.02cm of hy4]{$(0,\,1)$};
\node[bit, below=0.02cm of hy5]{$(1,\,0)$};
\node[bit, below=0.02cm of hy6]{$(1,\,1)$};
\node[bit, below=0.02cm of hy7]{$(0,\,1)$};
\node[bit, below=0.02cm of hy8]{$(1,\,0)$};

\node[lbl] at (0,-1.8) (Lz){Noise effect sequence\\$\mathbf{z}_4$};

\node[mac, right=0.2cm of Lz]  (z1){$0$};
\node[mac, right=0.08cm of z1] (z2){$0$};
\node[noi, right=0.08cm of z2] (z3){$1$};
\node[mac, right=0.08cm of z3] (z4){$0$};
\node[mac, right=0.08cm of z4] (z5){$0$};
\node[mac, right=0.08cm of z5] (z6){$0$};
\node[mac, right=0.08cm of z6] (z7){$0$};
\node[mac, right=0.08cm of z7] (z8){$0$};

\node[below=0.06cm of z1, font=\sffamily\tiny, text=black!40]{$k{=}1$};
\node[below=0.06cm of z2, font=\sffamily\tiny, text=black!40]{$k{=}2$};
\node[below=0.06cm of z3, font=\sffamily\tiny, text=red!65!black]{$k{=}3$};
\node[below=0.06cm of z4, font=\sffamily\tiny, text=black!40]{$k{=}4$};
\node[below=0.06cm of z5, font=\sffamily\tiny, text=black!40]{$k{=}5$};
\node[below=0.06cm of z6, font=\sffamily\tiny, text=black!40]{$k{=}6$};
\node[below=0.06cm of z7, font=\sffamily\tiny, text=black!40]{$k{=}7$};
\node[below=0.06cm of z8, font=\sffamily\tiny, text=black!40]{$k{=}8$};

\draw[swp] (z3.north) -- (hy3bit.south)
    node[midway, right=3pt, font=\sffamily\scriptsize, text=red!75!black]
    {substitute};

\node[lbl] at (0,-3.6) (Lty)
    {Substitution};

\node[mac,     right=0.2cm of Lty]  (ty1){$a_1$};
\node[mac,     right=0.08cm of ty1] (ty2){$a_2$};
\node[macswap, right=0.08cm of ty2] (ty3){$a_2$};
\node[mac,     right=0.08cm of ty3] (ty4){$a_3$};
\node[mac,     right=0.08cm of ty4] (ty5){$a_2$};
\node[mac,     right=0.08cm of ty5] (ty6){$a_4$};
\node[mac,     right=0.08cm of ty6] (ty7){$a_3$};
\node[mac,     right=0.08cm of ty7] (ty8){$a_2$};

\node[above=0.08cm of ty3, font=\sffamily\tiny, text=red!65!black]{\textbf{macrosymbol swap}};

\node[bit, below=0.02cm of ty1]{$(0,\,0)$};
\node[bit, below=0.02cm of ty2]{$(1,\,0)$};
\node[bit, below=0.02cm of ty3]{\textcolor{red!70!black}{$(1,\,0)$}};
\node[bit, below=0.02cm of ty4]{$(0,\,1)$};
\node[bit, below=0.02cm of ty5]{$(1,\,0)$};
\node[bit, below=0.02cm of ty6]{$(1,\,1)$};
\node[bit, below=0.02cm of ty7]{$(0,\,1)$};
\node[bit, below=0.02cm of ty8]{$(1,\,0)$};

\node[lbl] at (0,-5) (Lu1)
    {\textit{demod} User 1};

\node[bu1, right=0.2cm of Lu1]  (d11){$0$};
\node[bu1, right=0.08cm of d11] (d12){$1$};
\node[bu1, right=0.08cm of d12] (d13){\textbf{1}};
\node[bu1, right=0.08cm of d13] (d14){$0$};
\node[bu1, right=0.08cm of d14] (d15){$1$};
\node[bu1, right=0.08cm of d15] (d16){$1$};
\node[bu1, right=0.08cm of d16] (d17){$0$};
\node[bu1, right=0.08cm of d17] (d18){$1$};

\node[lbl] at (0,-5.8) (Lu2)
    {\textit{demod} User 2};

\node[bu2, right=0.2cm of Lu2]  (d21){$0$};
\node[bu2, right=0.08cm of d21] (d22){$0$};
\node[bu2, right=0.08cm of d22] (d23){$0$};
\node[bu2, right=0.08cm of d23] (d24){$1$};
\node[bu2, right=0.08cm of d24] (d25){$0$};
\node[bu2, right=0.08cm of d25] (d26){$1$};
\node[bu2, right=0.08cm of d26] (d27){$1$};
\node[bu2, right=0.08cm of d27] (d28){$0$};

\end{tikzpicture}
}
\caption{A macrosymbol swap for a particular $\mathbf{z}$ leaving User~2's bits unchanged. We consider the BPSK constellation
$\mathcal{X}^1 = \mathcal{X}^2 = \mathcal{X}$.
$\mathcal{X}_{\mathrm{agg}}$ is shown for a particular realization of $H^1, H^2$;
we let $\{a_1, a_2, a_3, a_4 \}\in \mathcal{X}_{\mathrm{agg}}$
for illustration. The per-user
bit pairs are shown beneath each macrosymbol.}
\label{fig:macroswap}
\vspace{-0.6cm}
\end{figure}